\documentclass{article}

\usepackage{arxiv}

\usepackage[utf8]{inputenc} 
\usepackage[T1]{fontenc}    
\usepackage{hyperref}       
\usepackage{url}            
\usepackage{booktabs}       
\usepackage{amsfonts}       
\usepackage{nicefrac}       
\usepackage{microtype}      
\usepackage{lipsum}         
\usepackage{graphicx}
\usepackage[sort&compress,square,numbers]{natbib}
\usepackage{doi}
\usepackage{authblk}           
\usepackage{stfloats} 
\usepackage{amsmath}
\usepackage{cleveref}       

\title{Nonlinear Inference Capacity of Fiber-Optical Extreme Learning Machines}

\author[1]{Sobhi Saeed}
\author[1]{Mehmet Müftüoğlu}
\author[1]{Glitta R. Cheeran} 
\author[1,2]{Thomas Bocklitz}
\author[1]{Bennet Fischer}
\author[1,3,*]{Mario Chemnitz}

\affil[1]{Leibniz-Institute of Photonic Technology, Albert-Einstein-Str. 9, 07745 Jena, Germany}
\affil[2]{Institute of Physical Chemistry, Friedrich Schiller University Jena, Helmholtzweg 4, 07743 Jena, Germany}
\affil[3]{Institute of Applied Optics and Biophysics, Friedrich Schiller University Jena, Albert-Einstein-Str. 15, 07745 Jena, Germany}
\affil[*]{mario.chemnitz@leibniz-ipht.de}

\date{\today} 

\begin{document}
\twocolumn[
\begin{@twocolumnfalse}
\maketitle

\begin{abstract}
The intrinsic complexity of nonlinear optical phenomena offers a fundamentally new resource to analog brain-inspired computing, with the potential to address the pressing energy requirements of artificial intelligence. We introduce and investigate the concept of nonlinear inference capacity in optical neuromorphic computing in highly nonlinear fiber-based optical Extreme Learning Machines. We demonstrate that this capacity scales with nonlinearity to the point where it surpasses the performance of a deep neural network model with five hidden layers on a scalable nonlinear classification benchmark. By comparing normal and anomalous dispersion fibers under various operating conditions and against digital classifiers, we observe a direct correlation between the system's nonlinear dynamics and its classification performance. Our findings suggest that image recognition tasks, such as MNIST, are incomplete in showcasing deep computing capabilities in analog hardware. Our approach provides a framework for evaluating and comparing computational capabilities, particularly their ability to emulate deep networks, across different physical and digital platforms, paving the way for a more generalized set of benchmarks for unconventional, physics-inspired computing architectures.
\end{abstract}
\end{@twocolumnfalse}
]

\twocolumn
\section{Introduction} 
\vspace*{-1pt}
The rapid advancement of artificial intelligence has sparked renewed interest in brain-inspired hardware, particularly optical implementations that promise energy-efficient solutions for AI acceleration and intelligent edge sensing. Traditional computing architectures face significant challenges when executing analog neural networks, resulting in substantial energy, water, and computational time requirements for operating large networks on conventional digital processors (i.e., CPUs, GPUs, TPUs).\\Optical approaches have garnered particular attention due to their intrinsic parallelism and scalability across multiple optical degrees of freedom, offering reduced energy consumption \cite{shekharRoadmappingNextGeneration2024}.
Unlike electronic solutions, a primary challenge in realizing competitive neuromorphic optical hardware lies in the all-optical implementation of nonlinearity to circumvent the electro-optical bottleneck toward deep all-optical architectures \cite{shastriSpatiotemporalPatternRecognition2014}, \cite{bandyopadhyaySinglechipPhotonicDeep2024}.\\ Nonlinearity is crucial for emulating synaptic switching behavior in information networks and enabling advanced learning capabilities, including improved accuracy and generalization. In deep networks, this issue is often addressed through multiple layers, which offer higher nonlinear mapping capabilities of the model. The concept of "nonlinear mapping capability" refers to a system's ability to transform input data into higher-dimensional spaces where previously inseparable patterns become linearly separable. 
Moving to hardware, physical substrates can inherently provide complex nonlinear responses through their natural dynamics without the need for additional layers or computational resources.
In optics, taking advantage of the nonlinearity offered by light-matter interactions in the media, while actively studied \cite{destrasSurveyActivationFunctions2024}, is widely assumed to be complicated and inefficient due to its high-power demands \cite{markovicPhysicsNeuromorphicComputing2020}.\\Several recent studies suggest alternatives to realize nontrivial nonlinearity in optical systems. These include electronic feedback loops to electro-optic modulators \cite{ortinUnifiedFrameworkReservoir2015}, \cite{Duport:12}, saturable absorption \cite{guptaMultilayerOpticalNeural2025} complex active gain dynamics in mode-competitive cavities \cite{skalliPhotonicNeuromorphicComputing2022}, and, most recently, repeated linear encoding through multiple scattering of input data \cite{wanjuraFullyNonlinearNeuromorphic2024}, \cite{xiaNonlinearOpticalEncoding2024}.\\
Neuromorphic computing with wave dynamics offers a promising approach to the elegant utilization of natural nonlinear dynamics in physical substrates. In optics, the concept aligns with Extreme Learning Machines (ELM) - a form of reservoir computing \cite{tanakaRecentAdvancesPhysical2019}, \cite{vandersandeAdvancesPhotonicReservoir2017} without internal recurrence. Unlike other optical implementations, the learning machine comprises multiple virtual nodes that are coupled by the intrinsic feed-forward propagation dynamics of a single optical component, such as a fiber. This concept, recently proposed by Marcucci et al. \cite{marcucciTheoryNeuromorphicComputing2020}, utilizes process-intrinsic, nonlinear mode interactions from natural waveguide dynamics as a computing resource. Experimental demonstrations include nonlinearly coupling spatial modes in multimode fibers \cite{teginScalableOpticalLearning2021} and spectral frequency generation in single-mode fibers \cite{Zhou2022a}, \cite{fischerNeuromorphicComputingFissionbased2023}, \cite{zajnulinaWeakKerrNonlinearity2023}.\\
However, analog wave computers have a common difficulty: the systems are generally complex and cannot be easily mapped to practical computing models, a critical challenge lies in the diversity of existing approaches and the general inability to measure a system's nonlinear mapping capabilities independent of specific tasks and to correlate them with learning abilities. This transformation is fundamental to solving complex tasks. Current approaches in the optical community empirically test new nonlinear systems on this capability through task-specific image recognition benchmarks. Here, it is common practice to compare, e.g., accuracies across system configurations to demonstrate improvements and to relate those improvements to the scaling in the learning behavior of deeper, hence more nonlinear neural networks computer models.\\
In particular, the MNIST dataset is frequently used as a benchmark. However, it is not highly nonlinear, as a logistic regression can achieve approximately 92\% accuracy \cite{palvanovComparisonsDeepLearning2018}. Similarly, linear Support Vector Machines (SVMs), which leverage kernel tricks to project input data into higher-dimensional spaces for linear separation, demonstrate comparable performance. Furthermore, performance improvements in big networks cannot be clearly attributed to stronger synaptic activation (i.e., deepness) yet might also result from increased connectivity as we will discuss further down in this paper. Task-specific benchmarks thus prove unsuitable for demonstrating deeper network activity. The question of nonlinearity's relevance in neural networks also remains largely unexplored in computer science, with only limited investigations into non-binary nonlinear activation functions, such as a multi-step perceptron \cite{huangExtremeLearningMachine2006} or trainable spline activation functions \cite{gaoLearningContinuousPiecewise2023}.\\
This work attempts to better understand and quantify the nonlinear inference capacity using a scalable, task-independent dataset and validate it using two different optical fiber processors. This investigation is particularly relevant for comparing computational capabilities across different physical substrates.\\ We begin with illustrating the operation principle of a frequency-based ELM in a single optical fiber, as an arbitrary kernel machine, we refer to it from this chapter onward as fiber-optical ELM. We then assess the inference capabilities of two different nonlinear systems distinguished by their dispersion properties and corresponding nonlinear dynamics (i.e., self-phase modulation and soliton fission).\\We assess the nonlinear inference capacity of the two different fiber-optical machines using a scalable spiral dataset, where increasing the angular span progressively challenges separability. 
These comparisons were conducted under different system configurations, including variations in fiber types, different numbers of system read-outs, and different power levels. Observations on both datasets are compared to digital classifier models, like neural networks and support vector machines, providing a strong basis for drawing parallels between the depth of digital networks and the nonlinear inference capacity of our analog computing systems.
\section{Materials and methods}
\subsection{Experimental setup}    
\vspace*{-1pt}
\begin{figure*}[b!]
    \centering
    \includegraphics[width=0.8\linewidth]{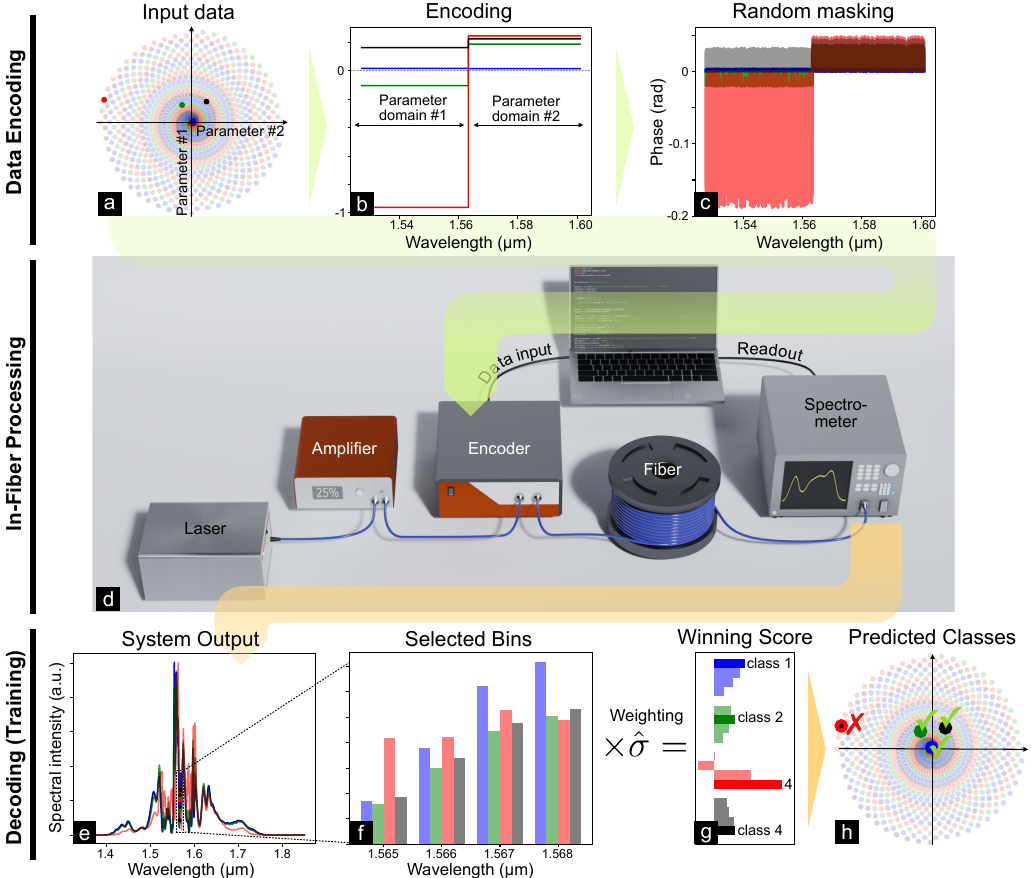}
    \caption{ Illustration of the data flow in the fiber-based neuromorphic system using an example from the spiral dataset. (a) Input data: four sample points were selected from four different spirals. (b) Corresponding data encoding: the first half of the spectral encoding range (limited by the WaveShaper) encodes the $X_1$-coordinate of a data tuple; the second half encodes the $X_2$-coordinate. (c) Encoding phase after multiplying with a constant phase scale factor and an arbitrary but fixed mask. (d) The experimental setup used for processing. A computer is used as I/O device and is not part of a feedback loop. (e) Linear spectral intensities at fiber output corresponding to the four sample inputs. (f) Linear spectral intensities at the selected, optimized search bins serving as system read-outs. (g) Prediction scores obtained by multiplying the read-outs with the trained weight matrix. Per sample, scores are sorted from class 1 to 4, from top to bottom. The highest values in a vector of four (i.e., argmax($\textbf{Y}^{score}$)) determines the predicted class. (h) Prediction results: points represent the predictions, while circles around these points indicate the true class labels, the examples contain one misclassification indicated by the red cross.}
    \label{fig:1}
\end{figure*}
We conducted experiments using a fiber-optical extreme learning machine in the frequency domain. The experimental system (see Fig. \ref{fig:1}d) comprises a femtosecond laser source (Toptica DFC) operating at 80 MHz repetition rate with ~70 nm bandwidth centered at 1560 nm, followed by an erbium-doped fiber amplifier (Thorlabs EDFA300p) operated at 25\% pump current to compensate for system losses. Data is encoded using a polarization-maintaining programmable spectral filter (Coherent Waveshaper 1000A/X) operating in the extended C- to L-band (1528-1602nm), which allows to independently modify spectral phase and amplitude across 400 frequency channels for both dispersion compensation and phase information imprinting. The initially chirped pulses are pre-compressed through a dispersion-compensating fiber (Thorlabs PMDCF, ~1m length) to 400 fs before entering the spectral filter. A particle swarm optimization algorithm in conjunction with optical autocorrelation measurements has been utilized to further optimize the input pulse phase to obtain 135 fs \cite{Efimov:98}, \cite{10.1117/12.3022291} before entering the main processing unit - a highly nonlinear fiber. Two nonlinear fibers were in use: (1) Thorlabs HN1550, 5m length with all-normal dispersion (-1 ps/nm/km @1550nm), and 10.8 $W^{-1}km^{-1}$ nonlinear coefficient, and (2) Thorlabs PMHN1, 5m length, with anomalous dispersion (+1 ps/nm/km @1550nm), and 10.7 $W^{-1}km^{-1}$ nonlinear coefficient.\\
The output is measured using an optical spectral analyzer (Yokogawa AQ6375E) and analyzed on a standard office computer. All components are fiber-coupled and polarization-maintaining, ensuring stable operation in a compact footprint.\\
We illustrate how the fiber-optical ELM processes information with the example of a four-arm spiral dataset. For the spiral classification benchmark, the inputs consist solely of two coordinates for each data point, with each coordinate tuple assigned to one point of the four spirals.\\Further details about the dataset are provided in the results section. The coordinate tuples are encoded into the spectral phase of a femtosecond pulse using a phase mask, where the first half of the phase mask corresponds to the $X_1$-coordinate and the second half corresponds to the $X_2$-coordinate, as illustrated in Fig. \ref{fig:1}b. To enhance the system's sensitivity to the encoded inputs, the phase mask is multiplied by a random mask, shown in Fig. \ref{fig:1}c. This approach is common for optical reservoir computers \cite{appeltantInformationProcessingUsing2011}, \cite{largerPhotonicInformationProcessing2012}, \cite{paquotOptoelectronicReservoirComputing2012}, \cite{martinenghiPhotonicNonlinearTransient2012}, as it leverages the system's sensitivity to phase jumps rather than absolute phase, ensuring a maximized system response. As a result, the output signals corresponding to different encodings become more distinguishable.
\subsection{Supervised offline training}    
\vspace*{-1pt}
Following the principles of extreme learning machines \cite{huangExtremeLearningMachine2006}, we train only the output layer. This method leverages the fiber-optical ELM's inherent ability to perform complex, high-dimensional transformations efficiently, reducing the computational burden associated with traditional training. 
Hence, to translate the output spectra of the system into an interpretable inference, we train a weight matrix based on selecting a subset from all wavelength recordings. From this chapter onward, we call the locations of the wavelength windows “search bins“ and their corresponding intensities “read-outs”.\\We select a user-defined number of search bins n using an optimization algorithm called  \textit{Equal Search}, a straightforward method that identifies combinations of equally spaced search bins, as described in \cite{fischerNeuromorphicComputingFissionbased2023}. From the resulting combinations, we choose the one that minimizes the mean square error $(MSE = \frac{1}{N} * \sum(\textbf{WX}-\textbf{Y})^2, N$ is the number of test data while applying cross validations) over five cross-validations. The weight matrix W ($R^{m} \times R^{n}$ with $m$ output classes and $n$ search bins) is calculated using solely linear regression, as given by the Moore-Penrose inverse 
\begin{equation}
\textbf{W}=(\textbf{X}^T \textbf{X})^{-1} \textbf{X}^T \textbf{Y}
\label{eq:1}
\end{equation}
where \textbf{X} is the read-outs vectors (intensities at selected search bins of dimension $R^n \times R^k$ with k samples), and \textbf{Y} is the task’s label vectors (ground truth of dimension $R^{m} \times R^{k}$). We experimented with incorporating the ridge term while training the system; however, we found it consistently yielded worse results (e.g., see long-term measurements in Fig. \ref{fig:2A} in appendix).\\
After training the weight matrix, classification is performed by acquiring the output values at the best combination of search bins identified before. These values are stored in a vector and multiplied by the weight matrix, producing another vector, referred to as the prediction score $\textbf{Y}^{score}=\textbf{WX}$. A winner-takes-all decision, i.e. $\textbf{Y}^{class} = \arg\max (\textbf{Y}^{score})$ is then performed on the Prediction Score vector, classifying the test data based on the index of the highest score, as shown in Figs. \ref{fig:1}g-h. We compare systems based on task-specific classification accuracy in percent ($\%$), which is defined by the ratio of number of correct class predictions over the total number of samples.\\

\begin{figure*}[!t]
    \centering
    \includegraphics[width=0.8\linewidth]{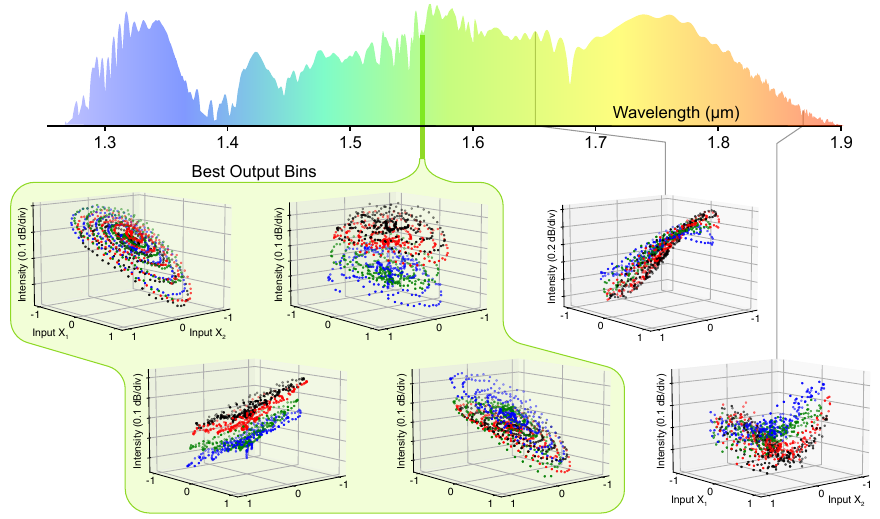}
    \caption{3D plots illustrating the relationship between output spectral intensity of a supercontinuum from an anomalous dispersive fiber and the input coordinates for all given samples. (a-d) Logarithmic spectral intensity versus input coordinates (X1, X2) at the optimized, selected search bins, demonstrating the system's intrinsically distinguishable response to different classes. (e-f) Spectral intensity versus input coordinates at two randomly selected wavelength windows. Both examples are still >10dB above the spectrometer’s noise floor.}
    \label{fig:2}
\end{figure*}

Fig. \ref{fig:2} illustrates the nonlinear projection capabilities of our fiber-optical neural network through measured spectral response to input data tuples ($X_1$, $X_2$) sampled from four distinct classes in a two-dimensional coordinate system. The \textit{Equal Search} algorithm identified four optimal search bins (Figs. \ref{fig:2}a->d) demonstrating highly distinct separation between the data classes in the spectral domain. Particularly noteworthy are the projections shown in Figs. \ref{fig:2}b-c, where the nonlinear fiber transformation achieves nearly ideal plane-wise separation of the classes in three dimensions, providing clear evidence of the system's powerful separation capabilities. In contrast, Figs. \ref{fig:2}e-f present randomly selected wavelength windows where the class separation is poor or non-existent, with data points from different classes showing significant overlap. 
This behavior effectively showcases the fundamental principle proposed by Huang et al. in their original ELM framework: ELMs can be understood as arbitrary kernel machines that, in optimal cases, provide hidden nodes with kernel response functions capable of transforming complex problems into linearly separable ones \cite{huangExtremeLearningMachine2006}. Following this interpretation, the performance of an ELM improves with access to a greater variety of kernel functions - a key insight that leads to our hypothesis that physical systems exhibiting richer nonlinear dynamics should outperform those with weaker nonlinearity in classification tasks.
\section{Results and discussion}
\vspace*{-1pt}
In this section, we demonstrate the capability of a fiber-optical ELM to effectively handle highly nonlinear tasks, exemplified by the spiral classification benchmark. The spiral classification task has emerged as a critical benchmark in the machine learning community \cite{changRobustPathbasedSpectral2008} for evaluating neural networks' capability to handle tangled data points and to learn nonlinear decision boundaries. This two-dimensional dataset, featuring interleaved spiral arms over an angular span of $2\pi$, is intrinsically not linearly separable and poses a particular challenge for optical computing systems due to the inherent difficulty in implementing nonlinear activation functions in the optical domain. However, recent breakthroughs have demonstrated significant progress in this area, with recent notable achievements including on-chip complex-valued neural networks reaching 95\% accuracy on a 2-arm spiral classification \cite{zhangOpticalNeuralChip2021} and a three-layer optoelectronic neural network achieving 86\% accuracy on a 4-arm spiral classification \cite{songLowpowerScalableMultilayer2024}.\\
Building upon these advances, we introduce an enhanced version of this benchmark that increases the hardness of the task. We generalized the spiral dataset by a free parameter, called the maximum angular span $\theta_{max}$, which controls the complexity of the spiral benchmark, making it particularly effective for illustrating the impact of the network depth and the system's nonlinear dynamics on the inference capabilities of digital and optical neuromorphic systems, respectively. As later shown in Figs. \ref{fig:5}a, \ref{fig:6}. As such, it is an excellent general measure of the nonlinear inference capacity of new neuromorphic physical substrates.\\
The data points are generated using analytic equations (Eq. 2), allowing flexibility in defining the number of phase-shifted spiral arms (number of classes, controlled by index $k$) and their wraps around the center (controls the complexity of the task). The number of wraps defined by a full 360° turn around the center is controlled by $\theta_{max}$, where the number of wraps equals  $\theta_{max}/2\pi$. For the presented results, we generated 250 points per spiral, yielding 1000 total points across four spirals. From this set we use 800 randomly selected points (200 per spiral) to train the system and kept 200 points (50 points per spiral) for testing after training (4:1 data split ratio).
\begin{equation}
\begin{aligned}
X_1 & =\frac{\theta}{\theta_{max}}\cos(\theta+k\frac{\pi}{2}) \quad \theta \in [0,\theta_{max}] \\
X_2 & =\frac{\theta}{\theta_{max}}\sin(\theta+k\frac{\pi}{2}) \quad k \in \{0,1,2,3\}
\end{aligned}
\label{eq:2}
\end{equation}
\begin{figure*}[!t]
    \centering
    \includegraphics[width=0.9\linewidth]{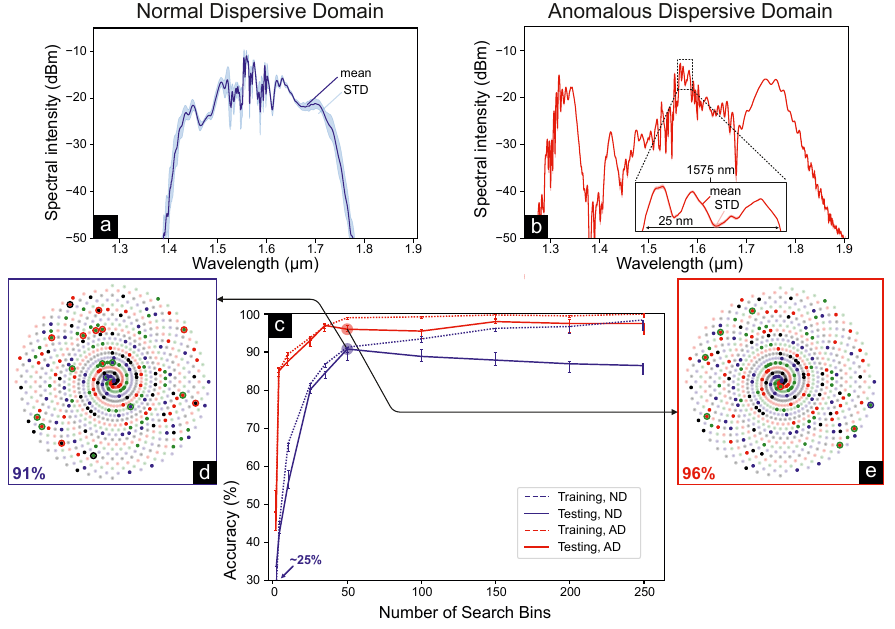}
    \caption{(a) Average and standard deviation (STD) of output spectral intensities for the spirals dataset in the normal dispersion (ND) case. (b) Average and std of output spectral intensities for the spirals dataset in the anomalous dispersion (AD) case. (c) Classification accuracy achieved for both fiber types as a function of the number of search bins. (d) Classification results for the ND case using 50 search bins. (e) Classification results for the AD case using 50 search bins.}
    \label{fig:3}
\end{figure*}
We evaluated our fiber-optical ELM on the spirals benchmark, with its increasing nonlinearity as $\theta_{max}$ increases (e.g. from $0.5\pi \ \ to \ 10\pi$), which is an ideal task to demonstrate the fiber-optical ELM’s ability to handle highly nonlinear problems. Fig. \ref{fig:3} illustrates the classification results for $\theta_{max}=10\pi$, corresponding to 5 wraps around the center per spiral. Figs. \ref{fig:3}a-b show the average and the standard deviation (STD) of the spectral intensities for both fiber types, normal dispersion (ND) fiber and anomalous dispersion (AD) fiber, respectively, measured under the same excitation conditions (i.e., 18.45 mW, 136 fs, cp. Tab. \ref{tab:1}).
The difference in standard deviations between the two cases arises from using different scale factors in the encoding phase for the two fibers.\\
We chose the scale factors according to the highest classification accuracy obtained during our experiments with various scale factors (see Fig. \ref{fig:4A} in Appendix). Specifically, we use $\pi/2$ for the ND fiber and $\pi/16$ for the AD fiber, resulting in a larger STD for the ND case. 
Our Equal Search analysis allows us to compare the system’s performance for various numbers of search bins. In Fig. \ref{fig:3}c, we observe increasing the number of search bins increases system accuracy for both fibers. The AD fiber achieves superior performance on the spiral benchmark (>95\% accuracy for 35 bins and up) compared to ND fiber, which reaches the best accuracy of 91.5\% with 50 bins before going into overfitting (i.e., test accuracy diverges from training accuracy).\\
\begin{figure*}[!b]
    \centering
    \includegraphics[width=1\linewidth]{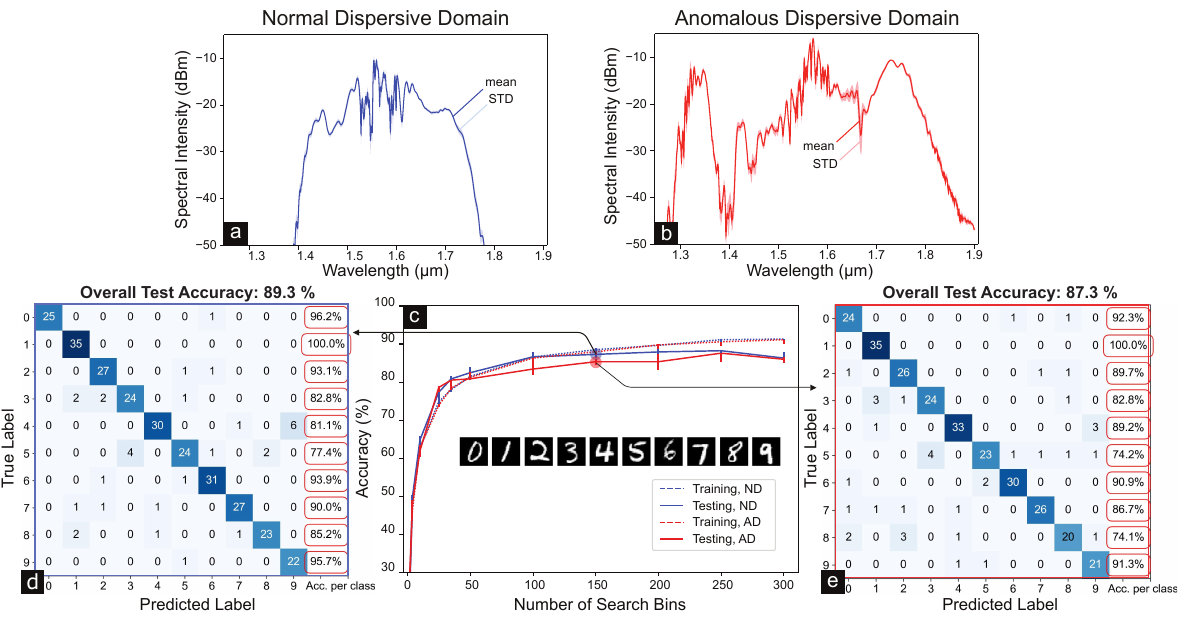}
    \caption{(a, b) Average and standard deviation (STD) of measured output spectral intensities for the MNIST dataset in the (a) normal dispersion (ND) case and (b) anomalous dispersion (AD) case. (c) Classification accuracy achieved across 300 MNIST test samples as a function of the number of search bins for both fiber types. (d, e) Confusion matrices of our systems for unseen test data for the (d) ND case (achieved accuracy 89.33\%), and (e) AD case (achieved accuracy 87.3\%) using 150 search bins for both cases.}
    \label{fig:4}
\end{figure*}
We test the system’s performance on a low nonlinear task known to be solvable to a high degree by simply using linear classifiers. We identified this dataset to be suitable for this purpose. The MNIST dataset is a widely used collection of handwritten digits from 0 to 9 (i.e., 10 classes), stored as $28\times28$ pixels gray-valued images, that serves as a standard benchmark in machine learning. The dataset used in this study is sourced from the TensorFlow library (tensorflow$\_$datasets).\\
We trained the system on the first 2100 images and tested it on 300 images previously unseen by the system using both normal and anomalous dispersion fibers. To encode the MNIST images, we first crop an 18×18-pixel window to meet the data input limitations of the Waveshaper. The cropped images were then flattened row-wise from top to bottom into 1D signals corresponding to pixel intensities. These 1D signals were multiplied by a random mask and a phase scale factor before adding them to an optimized phase profile, yielding the encoding phase (cp. Fig. \ref{fig:3A}d) which modifies the optical signal using the Waveshaper, hence encoding the input information, before propagating through the fiber. At the output layer, the system follows the same procedure as in the spiral benchmark. It stores the output read-outs at the newly selected search bins and translates these read-outs into an interpretable prediction score by multiplying the read-outs vector with a trained weight matrix. The highest value in the prediction score (i.e., argmax($\textbf{Y}^{score}$)) determines the predicted class. Fig. \ref{fig:3A} in Appendix illustrates all input and output levels for an MNIST test image.\\
Unlike the spiral benchmark, the system demonstrates similar performance on MNIST when using ND fiber (89.33\% accuracy) and AD fiber (87.3\% accuracy), as shown in Figs. \ref{fig:4}d-e, respectively. Both cases surpass the baseline of 83.7\% for linear regression on the data \cite{haryPrinciplesMetricsExtreme2025} and perform similarly to a support vector machine (SVM) with a linear kernel which achieves about 91\% in accuracy with only 20 support vectors (cp. Figure 1A in Appendix). Classification accuracy further improves with an increased number of search bins, as illustrated in Fig. \ref{fig:4}c.\\
We hypothesize that the higher nonlinear mapping introduced by the AD fiber does not significantly affect system performance for a task with low nonlinearity, such as MNIST. The slightly better performance observed with the ND fiber in our setup can be attributed to a more extensive optimization process during the selection of the optimal encoding phase scale factor. Specifically, when using the ND fiber, we conducted multiple experiments with different phase scale factors and selected the one that achieved the highest accuracy. For both fibers, a phase scale factor of $\frac{\pi}{4}$ was ultimately chosen.
Figs. \ref{fig:4}a-b depict the average spectral output intensities and their corresponding standard deviations for both ND and AD fibers. The STD values in the two cases are comparable because we used the same phase scale factor.\\
The confusion matrices and per-class accuracy for both cases are presented in Figs. \ref{fig:4}d-e. A notable observation is the misclassification between the digit classes  \verb!3! and \verb!5! and classes \verb!4! and \verb!9!, which likely arises from their similar 1D intensity profiles. \\
To assess robustness, we conducted a long-term evaluation involving training the system on 3000 images and testing it on 35000 images  over an acquisition period of 35 hours. The system maintained an accuracy of 84.69\%, with only a modest 4.5\% drop over the extended testing period, underscoring its reliability and resilience. Further details of this robustness analysis are provided in Fig. \ref{fig:2A} in the appendix.\\
We further investigate the impact of input power (consequently the system's nonlinear mapping capability) on the performance of our fiber-optical ELM handling increasingly harder task, by changing the attenuation applied on the input signals (0, 2, 4, 6, 8 dB), while progressively increasing the spirals' $\theta_{max} (0.5\pi,\ 2\pi,\ 4\pi,\ 8\pi,\ 10\pi)$, thereby increasing the complexity of the classification task. 
Additionally, we benchmark our system against digital neural networks with different configurations for both the spiral and MNIST datasets. We based our comparison between the optical and digital systems solely on the achieved results as we fed both systems with the exact same dataset with a fixed training-testing split ratio.\\
The results highlight the fiber-optical ELM’s efficiency in handling highly nonlinear tasks, such as the spiral benchmark with a large angular span ($\theta_{max}$), outperforming a 5-hidden-layer digital neural network with 1024 nodes for $\theta_{max} > 10\pi $. Furthermore, the results demonstrate the critical role of network depth to address highly nonlinear tasks in digital neural networks. As shown in Fig. \ref{fig:5}a, deeper networks offer significant performance advantages when handling tasks with high nonlinearity, despite having the same number of nodes as their shallower counterparts. Distribution of the nodes is illustrated in Tab. \ref{tab:1A} in Appendix.
Fig. \ref{fig:5}b illustrates the accuracy achieved using the ND fiber on the spiral benchmark with different maximum angular spans $(\theta_{max}=\ 0.5\pi,\ 2\pi,\ 4\pi,\ 8\pi,\ 10\pi)$, under different attenuation levels (0, 2, 4, 6, 8 dB), represented as heatmap rows and columns, respectively. Fig. \ref{fig:5}c presents the corresponding results for the AD fiber. Both cases use 50 search bins. We notice that applying attenuation (decreasing the input power) has a more pronounced effect on system performance for highly nonlinear tasks (e.g., spirals with high $\theta_{max}$) compared to its impact on low nonlinear tasks (spirals with low $\theta_{max}$ and MNIST, cp. Fig. \ref{fig:6}). This performance degradation stems from the reduced nonlinear mapping capability of the system at lower input power levels. However, increasing the number of search bins mitigates the loss in nonlinear mapping induced by attenuation (cp. Fig. \ref{fig:6}a), this is in accordance with previous observations on comparable problems \cite{fischerNeuromorphicComputingFissionbased2023}. 
Interestingly, the AD fiber shows particularly better performance at lower input power levels for more complex tasks compared the ND fiber, though with greater fluctuations. These fluctuations are most evident in the unexpectedly high accuracy for $\theta_{max}=\ 8\pi$ at attenuation of 6 dB (input power = 6.7 dBm). 
In all subfigures of Fig. \ref{fig:5}, the system nonlinearity is defined by the soliton number:
\begin{equation}
    N=\ \sqrt{\frac{\gamma P_0T_0^2}{\left|\beta_2\right|}}
    \label{eq:3}
\end{equation}
\begin{equation}
    P_0=\frac{P_{avg}}{T_0 \cdot f_{rep}} 
    \label{eq:4}
\end{equation}
$\gamma$ is the nonlinear coefficient $(W^{-1} km^{-1})$, $P_0$ is the peak power, $T_0$ is the measured pulse width (seconds)  and $\beta_2$ is the dispersion coefficient $({s^2}/{m})$, and $f_{rep}$ is the laser repetition rate in (Hz). Tab. \ref{tab:1} shows the power and autocorrelation measurements for the presented results.\\
\begin{figure}
    \centering
    \includegraphics[width=1\linewidth]{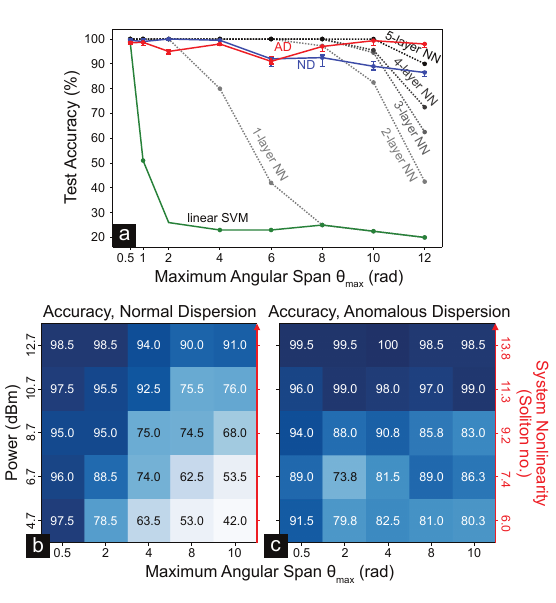}
    \caption{(a) Best test accuracy on 200 spiral data samples achieved by digital classifiers (a linear kernel support vector machine with 100 support vectors, and neural networks in different configurations (cp. Tab. \ref{tab:1A}); all trained for 1000 training epochs) and our fiber-optical ELM using 100 search bins for increasing nonlinear problem hardness in the spiral task, defined by the maximum angular span $\theta_{max}$. (b,c) Test accuracies on 200 spiral data samples as a function of system nonlinearity (or attenuation) and maximum angular span for both, (b) normal dispersion and (c) anomalous dispersion.}
    \label{fig:5}
\end{figure}
\begin{table}[!htbp]
\centering
\caption{Pulse width and peak power at the input of the highly nonlinear fiber (HNLF) for various encoding conditions (attenuation), calculated using average power and pulse width (both are measured before propagating through the fiber). The pulse width was measured using an APE autocorrelator (pulseCheck NX 50). The average of 16 scans over a range of 5 ps was fitted to a Lorentzian function to determine the pulse width.}
\label{tab:1}
\begin{tabular}{p{1.7cm}|p{1.7cm}|p{1.7cm}|p{1.7cm}}
\textbf{Encoding attenuation applied (dB)} & \textbf{Average power measured (mW)} & \textbf{Pulse width measured (fs)} & \textbf{Peak power calculated (W)} \\ \hline
0 & 18.56 & 136 & 1705.88 \\ 
2 & 11.66 & 146 & 998.29 \\ 
4 & 7.41 & 155 & 597.58 \\ 
6 & 4.72 & 157.5 & 374.60 \\ 
8 & 3.04 & 161 & 236.02 \\ 
\end{tabular}
\end{table}
The soliton number corresponds to a normalized nonlinear gain parameter of a waveguide system which is proportional with peak power $P_0$ and, hence, inversely to the attenuation level. We emphasize parametrizing the system's nonlinear mapping capability using the soliton number rather than input power alone, as it reflects both the peak power and the pulse width of the input signal, hence the signal’s coherence, which is essential for supercontinuum generation \cite{Castello-Lurbe:20}.\\
Lastly, we demonstrate that increasing the optical system's nonlinearity (indicated by the soliton number) has a similar impact on the achieved results in an optical neuromorphic system, to increasing the depth of a digital neural network. Particularly, investigating these performance trends similarities on a highly nonlinear task represented by the spiral benchmark with an angular span of $\theta_{max}=10\pi$ and a low nonlinear task represented by MNIST.\\
Fig. \ref{fig:6}a presents the test accuracy achieved for MNIST and the spiral benchmark (with $\theta_{max}=10\pi$) at various attenuation levels. The results are shown for two configurations: using 150 search bins (solid lines) and 50 search bins (dotted lines), while coupling with an ND fiber and maintaining the same system configuration described earlier.\\
We observe that increasing the system’s nonlinearity does not enhance its performance on the MNIST dataset. In contrast, higher nonlinearity significantly improves performance on the spiral benchmark with a high angular span ($\theta_{max}=10\pi$). A similar trend is evident when solving these benchmarks using digital neural networks. Fig. 6b illustrates the achieved test accuracy for different configurations of a multilayer neural network, yet they all have the same total number of nodes, 400.\\ MNIST is solvable with 10 nodes as shown in Fig. \ref{fig:1A} in Appendix. However, we needed a bigger network to solve the spiral Benchmark. The 400 nodes are distributed across the hidden layers as shown in Tab. \ref{tab: 2A}. Number of hidden layers is indicated by the x axis in Fig. \ref{fig:6}b.
From Fig. \ref{fig:6}b, as expected, increasing the depth of the network has a negligible impact on the test accuracy achieved for the MNIST dataset, they only demonstrate faster convergence, enabling the model to stabilize with fewer training epochs. In contrast, deeper networks provide a significant advantage when handling highly nonlinear tasks, such as the spiral benchmark with $\theta_{max} = 10\pi$. Specifically, 4- and 5-hidden-layer networks with 400 nodes (see Fig. \ref{fig:6}b) outperform 2- and 3-hidden-layer networks with 1024 nodes (see Fig. \ref{fig:5}a), highlighting the importance of networks depth when handling complex nonlinear tasks.
\begin{figure}[!t]
    \centering
    \includegraphics[width=\linewidth]{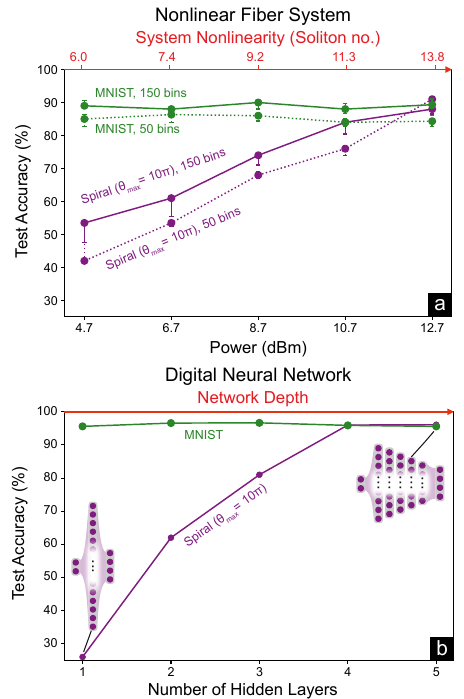}
    \caption{(a) Performance trends of the fiber-optical ELM on a highly nonlinear task (spiral benchmark) and a low nonlinear task (MNIST) under varying attenuation levels. (b) Performance trends of a multi-layer neural network with different numbers of hidden layers but the same total number of nodes (400), evaluated on the spiral benchmark and MNIST datasets.}
    \label{fig:6}
\end{figure}
\section{Conclusion}
Our results suggest that while the performance of fiber-optical ELMs stalls on the MNIST dataset, they excel in handling highly nonlinear tasks. This highlights their suitability for specialized applications where conventional optical neural networks, which are often shallow (i.e., single-layer), face limitations.\\
To date, the performance gain in optical neural networks is traditionally shown in improvements in image classification results across various datasets, including MNIST, fashion-MNIST, ImageNet. While these datasets are widely used, they fail to isolate the performance gains achieved through nonlinearity from those resulting from increased network connectivity. We demonstrated that measuring the system’s nonlinear inference capacity through scalable nonlinear benchmarks, such as the spiral dataset, provides a less ambiguous way to compare the computational depth of unconventional neuromorphic hardware. The spiral benchmark's key strength lies in its parametric nature, allowing controlled increments in task complexity that directly test a system's nonlinear inference capacity. This approach has proven especially valuable in two critical aspects: First, it enables quantitative comparison between different analog computing substrates, whose computational operations are often difficult to enumerate in the number of performed computes. Second, it establishes a direct bridge between physical implementations and digital models, allowing us to correlate the scaling of physical system dynamics with the depth of digital neural networks. These findings may pave the way for a framework for developing standardized benchmarks that can effectively evaluate both conventional and physics-inspired computing architectures, particularly in their ability to handle increasingly complex nonlinear tasks.\\
We furthermore observe a direct separation of the spiral classes in selected output channels of our systems, which uniquely showcase nonlinear Schrödinger systems as genuinely arbitrary kernel machines, providing a unique substrate for implementing the ELM computing framework. The soliton number serves as a valuable, platform-independent measure of the nonlinear inference capacity. By incorporating this parameter, we gain a more nuanced understanding of the system’s behavior and its ability to process complex inputs effectively.\\
Through our work, we hope to encourage ongoing exploration of generalizable performance metrics, as the connection between data-driven inference performance and the measures of intricate physical dynamics may prove vital in uncovering the genuine computational performance of natural systems. Information measures, such as Shannon Entropy \cite{zajnulinaShannonEntropyHelps2024}, Fischer Information \cite{rotterConceptFisherInformation2024}, or Information Processing Capacity \cite{dambre2012information} may provide a profound start, yet their application to analog feedforward ELMs remains unclear.\\
Our findings collectively highlight the promise of the fiber-optical Schrödinger system as a scalable, efficient solution for neuromorphic computing. Their capability to handle highly nonlinear tasks with a compact and stable design positions them as a strong candidate for advancing future computing systems, in particular in a more open approach to unconventional neuromorphic computing frameworks that go beyond the perceptron architecture.\\
\textbf{Acknowledgements}: We acknowledge that this work was made possible by funding from the Carl Zeiss Foundation through the NEXUS program (project P2021-05-025).\\
\textbf{Research Funding}: Carl Zeiss Stiftung\\
\textbf{Author Contributions}: S.S. did the data acquisition, data analysis, modeling, and coding, and contributed to writing, proof reading, figure editing, and methods development. M.C. led the project, supervised the data acquisition, data analysis, modeling, and coding, and contributed to writing, proof reading, figure editing, and methods development. M.C. and B.F. designed the setup. B.F also contributed to optimizing the experiments, coding, proof reading, methods development and figure editing. M.M. contributed to optimizing the experiments, coding, proof reading. G.R.C. contributed to data analysis, coding, and proof reading. T.B. contributed to modeling and proof reading.\\
\textbf{Conflict of Interest Statement}: The authors declare no conflict of interest regarding the publication of this paper.\\
\textbf{Data Availability}: The used raw spiral benchmark data is available on our Github page. \url{DOI: 10.5281/zenodo.14773759}.\\ 

\bibliographystyle{unsrtnat}
\bibliography{References} 

\begin{thebibliography}{35}
\providecommand{\natexlab}[1]{#1}
\providecommand{\url}[1]{\texttt{#1}}
\expandafter\ifx\csname urlstyle\endcsname\relax
  \providecommand{\doi}[1]{doi: #1}\else
  \providecommand{\doi}{doi: \begingroup \urlstyle{rm}\Url}\fi

\bibitem[Shekhar et~al.()Shekhar, Bogaerts, Chrostowski, Bowers, Hochberg,
  Soref, and Shastri]{shekharRoadmappingNextGeneration2024}
Sudip Shekhar, Wim Bogaerts, Lukas Chrostowski, John~E. Bowers, Michael
  Hochberg, Richard Soref, and Bhavin~J. Shastri.
\newblock Roadmapping the next generation of silicon photonics.
\newblock 15\penalty0 (1):\penalty0 751.
\newblock ISSN 2041-1723.
\newblock \doi{10.1038/s41467-024-44750-0}.
\newblock URL \url{https://www.nature.com/articles/s41467-024-44750-0}.
\newblock Publisher: Nature Publishing Group.

\bibitem[Shastri et~al.()Shastri, Tait, Nahmias, {Ben Wu}, and
  Prucnal]{shastriSpatiotemporalPatternRecognition2014}
Bhavin~J. Shastri, Alexander~N. Tait, Mitchell~A. Nahmias, {Ben Wu}, and
  Paul~R. Prucnal.
\newblock Spatiotemporal pattern recognition with cascadable graphene excitable
  lasers.
\newblock In \emph{2014 {IEEE} Photonics Conference}, pages 573--574. {IEEE}.
\newblock ISBN 978-1-4577-1504-4.
\newblock \doi{10.1109/IPCon.2014.6995270}.
\newblock URL \url{http://ieeexplore.ieee.org/document/6995270/}.

\bibitem[Bandyopadhyay et~al.()Bandyopadhyay, Sludds, Krastanov, Hamerly,
  Harris, Bunandar, Streshinsky, Hochberg, and
  Englund]{bandyopadhyaySinglechipPhotonicDeep2024}
Saumil Bandyopadhyay, Alexander Sludds, Stefan Krastanov, Ryan Hamerly,
  Nicholas Harris, Darius Bunandar, Matthew Streshinsky, Michael Hochberg, and
  Dirk Englund.
\newblock Single-chip photonic deep neural network with forward-only training.
\newblock 18\penalty0 (12):\penalty0 1335--1343.
\newblock ISSN 1749-4885, 1749-4893.
\newblock \doi{10.1038/s41566-024-01567-z}.
\newblock URL \url{https://www.nature.com/articles/s41566-024-01567-z}.

\bibitem[Destras et~al.()Destras, Le~Beux, De~Magalhães, and
  Nicolescu]{destrasSurveyActivationFunctions2024}
Oceane Destras, Sébastien Le~Beux, Felipe~Gohring De~Magalhães, and Gabriela
  Nicolescu.
\newblock Survey on activation functions for optical neural networks.
\newblock 56\penalty0 (2):\penalty0 1--30.
\newblock ISSN 0360-0300, 1557-7341.
\newblock \doi{10.1145/3607533}.
\newblock URL \url{https://dl.acm.org/doi/10.1145/3607533}.

\bibitem[Marković et~al.()Marković, Mizrahi, Querlioz, and
  Grollier]{markovicPhysicsNeuromorphicComputing2020}
Danijela Marković, Alice Mizrahi, Damien Querlioz, and Julie Grollier.
\newblock Physics for neuromorphic computing.
\newblock 2\penalty0 (9):\penalty0 499--510.
\newblock ISSN 2522-5820.
\newblock \doi{10.1038/s42254-020-0208-2}.
\newblock URL \url{https://doi.org/10.1038/s42254-020-0208-2}.

\bibitem[Ortín et~al.()Ortín, Soriano, Pesquera, Brunner, San-Martín,
  Fischer, Mirasso, and Gutiérrez]{ortinUnifiedFrameworkReservoir2015}
S.~Ortín, M.~C. Soriano, L.~Pesquera, D.~Brunner, D.~San-Martín, I.~Fischer,
  C.~R. Mirasso, and J.~M. Gutiérrez.
\newblock A unified framework for reservoir computing and extreme learning
  machines based on a single time-delayed neuron.
\newblock 5\penalty0 (1):\penalty0 14945.
\newblock ISSN 2045-2322.
\newblock \doi{10.1038/srep14945}.
\newblock URL \url{http://www.nature.com/articles/srep14945}.

\bibitem[Duport et~al.(2012)Duport, Schneider, Smerieri, Haelterman, and
  Massar]{Duport:12}
Fran\c{c}ois Duport, Bendix Schneider, Anteo Smerieri, Marc Haelterman, and
  Serge Massar.
\newblock All-optical reservoir computing.
\newblock \emph{Opt. Express}, 20\penalty0 (20):\penalty0 22783--22795, Sep
  2012.
\newblock \doi{10.1364/OE.20.022783}.
\newblock URL \url{https://opg.optica.org/oe/abstract.cfm?URI=oe-20-20-22783}.

\bibitem[Gupta et~al.()Gupta, Lee, Cho, and
  Choi]{guptaMultilayerOpticalNeural2025}
Kalpak Gupta, Ye-Ryoung Lee, Ye-Chan Cho, and Wonshik Choi.
\newblock Multilayer optical neural network using saturable absorber for
  nonlinearity.
\newblock 577:\penalty0 131471.
\newblock ISSN 00304018.
\newblock \doi{10.1016/j.optcom.2024.131471}.
\newblock URL
  \url{https://linkinghub.elsevier.com/retrieve/pii/S0030401824012082}.

\bibitem[Skalli et~al.()Skalli, Robertson, Owen-Newns, Hejda, Porte,
  Reitzenstein, Hurtado, and Brunner]{skalliPhotonicNeuromorphicComputing2022}
Anas Skalli, Joshua Robertson, Dafydd Owen-Newns, Matej Hejda, Xavier Porte,
  Stephan Reitzenstein, Antonio Hurtado, and Daniel Brunner.
\newblock Photonic neuromorphic computing using vertical cavity semiconductor
  lasers.
\newblock 12\penalty0 (6):\penalty0 2395.
\newblock ISSN 2159-3930.
\newblock \doi{10.1364/OME.450926}.
\newblock URL \url{https://opg.optica.org/abstract.cfm?URI=ome-12-6-2395}.

\bibitem[Wanjura and Marquardt()]{wanjuraFullyNonlinearNeuromorphic2024}
Clara~C. Wanjura and Florian Marquardt.
\newblock Fully nonlinear neuromorphic computing with linear wave scattering.
\newblock 20\penalty0 (9):\penalty0 1434--1440.
\newblock ISSN 1745-2481.
\newblock \doi{10.1038/s41567-024-02534-9}.
\newblock URL \url{https://www.nature.com/articles/s41567-024-02534-9}.
\newblock Publisher: Nature Publishing Group.

\bibitem[Xia et~al.()Xia, Kim, Eliezer, Han, Shaughnessy, Gigan, and
  Cao]{xiaNonlinearOpticalEncoding2024}
Fei Xia, Kyungduk Kim, Yaniv Eliezer, {SeungYun} Han, Liam Shaughnessy, Sylvain
  Gigan, and Hui Cao.
\newblock Nonlinear optical encoding enabled by recurrent linear scattering.
\newblock pages 1--9.
\newblock ISSN 1749-4893.
\newblock \doi{10.1038/s41566-024-01493-0}.
\newblock URL \url{https://www.nature.com/articles/s41566-024-01493-0}.

\bibitem[Tanaka et~al.()Tanaka, Yamane, Héroux, Nakane, Kanazawa, Takeda,
  Numata, Nakano, and Hirose]{tanakaRecentAdvancesPhysical2019}
Gouhei Tanaka, Toshiyuki Yamane, Jean~Benoit Héroux, Ryosho Nakane, Naoki
  Kanazawa, Seiji Takeda, Hidetoshi Numata, Daiju Nakano, and Akira Hirose.
\newblock Recent advances in physical reservoir computing: A review.
\newblock 115:\penalty0 100--123.
\newblock ISSN 0893-6080.
\newblock \doi{10.1016/j.neunet.2019.03.005}.
\newblock URL
  \url{https://www.sciencedirect.com/science/article/pii/S0893608019300784}.

\bibitem[Van Der~Sande et~al.()Van Der~Sande, Brunner, and
  Soriano]{vandersandeAdvancesPhotonicReservoir2017}
Guy Van Der~Sande, Daniel Brunner, and Miguel~C. Soriano.
\newblock Advances in photonic reservoir computing.
\newblock 6\penalty0 (3):\penalty0 561--576.
\newblock ISSN 2192-8614, 2192-8606.
\newblock \doi{10.1515/nanoph-2016-0132}.
\newblock URL
  \url{https://www.degruyter.com/document/doi/10.1515/nanoph-2016-0132/html}.

\bibitem[Marcucci et~al.()Marcucci, Pierangeli, and
  Conti]{marcucciTheoryNeuromorphicComputing2020}
Giulia Marcucci, Davide Pierangeli, and Claudio Conti.
\newblock Theory of neuromorphic computing by waves: Machine learning by rogue
  waves, dispersive shocks, and solitons.
\newblock 125\penalty0 (9):\penalty0 093901.
\newblock ISSN 0031-9007, 1079-7114.
\newblock \doi{10.1103/PhysRevLett.125.093901}.
\newblock URL \url{https://link.aps.org/doi/10.1103/PhysRevLett.125.093901}.

\bibitem[Teğin et~al.()Teğin, Yıldırım, Oğuz, Moser, and
  Psaltis]{teginScalableOpticalLearning2021}
Uğur Teğin, Mustafa Yıldırım, İlker Oğuz, Christophe Moser, and Demetri
  Psaltis.
\newblock Scalable optical learning operator.
\newblock 1\penalty0 (8):\penalty0 542--549.
\newblock ISSN 2662-8457.
\newblock \doi{10.1038/s43588-021-00112-0}.
\newblock URL \url{https://www.nature.com/articles/s43588-021-00112-0}.

\bibitem[Zhou et~al.()Zhou, Scalzo, and Jalali]{Zhou2022a}
Tingyi Zhou, Fabien Scalzo, and Bahram Jalali.
\newblock Nonlinear schrödinger kernel for hardware acceleration of machine
  learning.
\newblock 40\penalty0 (5):\penalty0 1308--1319.
\newblock ISSN 0733-8724.
\newblock \doi{10.1109/JLT.2022.3146131}.
\newblock URL \url{https://ieeexplore.ieee.org/document/9695395/}.

\bibitem[Fischer et~al.()Fischer, Chemnitz, Zhu, Perron, Roztocki, {MacLellan},
  Di~Lauro, Aadhi, Rimoldi, Falk, and
  Morandotti]{fischerNeuromorphicComputingFissionbased2023}
Bennet Fischer, Mario Chemnitz, Yi~Zhu, Nicolas Perron, Piotr Roztocki,
  Benjamin {MacLellan}, Luigi Di~Lauro, A.~Aadhi, Cristina Rimoldi, Tiago~H.
  Falk, and Roberto Morandotti.
\newblock Neuromorphic computing via fission‐based broadband frequency
  generation.
\newblock 10\penalty0 (35):\penalty0 2303835.
\newblock ISSN 2198-3844, 2198-3844.
\newblock \doi{10.1002/advs.202303835}.
\newblock URL \url{https://onlinelibrary.wiley.com/doi/10.1002/advs.202303835}.

\bibitem[Zajnulina et~al.()Zajnulina, Lupo, and
  Massar]{zajnulinaWeakKerrNonlinearity2023}
Marina Zajnulina, Alessandro Lupo, and Serge Massar.
\newblock Weak kerr nonlinearity boosts the performance of
  frequency-multiplexed photonic extreme learning machines: A multifaceted
  approach.
\newblock URL \url{http://arxiv.org/abs/2312.12296}.

\bibitem[Palvanov and Cho()]{palvanovComparisonsDeepLearning2018}
Akmaljon Palvanov and Young~Im Cho.
\newblock Comparisons of deep learning algorithms for {MNIST} in real-time
  environment.
\newblock 18\penalty0 (2):\penalty0 126--134.
\newblock ISSN 1598-2645, 2093-744X.
\newblock \doi{10.5391/IJFIS.2018.18.2.126}.
\newblock URL
  \url{http://www.ijfis.org/journal/view.html?uid=824&sort=&scale=&key=year&keyword=&s_v=18&s_n=2&pn=vol&year=2018&vmd=Full}.

\bibitem[Huang et~al.()Huang, Zhu, and Siew]{huangExtremeLearningMachine2006}
Guang-Bin Huang, Qin-Yu Zhu, and Chee-Kheong Siew.
\newblock Extreme learning machine: Theory and applications.
\newblock 70\penalty0 (1):\penalty0 489--501.
\newblock ISSN 09252312.
\newblock \doi{10.1016/j.neucom.2005.12.126}.
\newblock URL
  \url{https://linkinghub.elsevier.com/retrieve/pii/S0925231206000385}.

\bibitem[Gao et~al.()Gao, Li, Li, Duan, Van~Gool, Benini, and
  Magno]{gaoLearningContinuousPiecewise2023}
Xinchen Gao, Yawei Li, Wen Li, Lixin Duan, Luc Van~Gool, Luca Benini, and
  Michele Magno.
\newblock Learning continuous piecewise non-linear activation functions for
  deep neural networks.
\newblock In \emph{2023 {IEEE} International Conference on Multimedia and Expo
  ({ICME})}, pages 1835--1840.
\newblock \doi{10.1109/ICME55011.2023.00315}.
\newblock URL \url{https://ieeexplore.ieee.org/document/10219752}.
\newblock {ISSN}: 1945-788X.

\bibitem[Efimov et~al.(1998)Efimov, Moores, Beach, Krause, and
  Reitze]{Efimov:98}
Anatoly Efimov, Mark~D. Moores, Nicole~M. Beach, Jeffrey~L. Krause, and
  David~H. Reitze.
\newblock Adaptive control of pulse phase in a chirped-pulse amplifier.
\newblock \emph{Opt. Lett.}, 23\penalty0 (24):\penalty0 1915--1917, Dec 1998.
\newblock \doi{10.1364/OL.23.001915}.
\newblock URL \url{https://opg.optica.org/ol/abstract.cfm?URI=ol-23-24-1915}.

\bibitem[Fischer et~al.(2024)Fischer, M{\"u}ft{\"u}oglu, and
  Chemnitz]{10.1117/12.3022291}
Bennet Fischer, Mehmet M{\"u}ft{\"u}oglu, and Mario Chemnitz.
\newblock {Machine-aided near-transform-limited pulse compression in fully
  fiber-interconnected systems for efficient spectral broadening}.
\newblock In John~M. Dudley, Anna~C. Peacock, Birgit Stiller, and Giovanna
  Tissoni, editors, \emph{Nonlinear Optics and its Applications 2024}, volume
  13004, page 1300402. International Society for Optics and Photonics, SPIE,
  2024.
\newblock \doi{10.1117/12.3022291}.
\newblock URL \url{https://doi.org/10.1117/12.3022291}.

\bibitem[Appeltant et~al.(2011)Appeltant, Soriano, Van Der~Sande, Danckaert,
  Massar, Dambre, Schrauwen, Mirasso, and
  Fischer]{appeltantInformationProcessingUsing2011}
L.~Appeltant, M.C. Soriano, G.~Van Der~Sande, J.~Danckaert, S.~Massar,
  J.~Dambre, B.~Schrauwen, C.R. Mirasso, and I.~Fischer.
\newblock Information processing using a single dynamical node as complex
  system.
\newblock \emph{Nature Communications}, 2\penalty0 (1):\penalty0 468, September
  2011.
\newblock ISSN 2041-1723.
\newblock \doi{10.1038/ncomms1476}.
\newblock URL \url{https://www.nature.com/articles/ncomms1476}.

\bibitem[Larger et~al.()Larger, Soriano, Brunner, Appeltant, Gutierrez,
  Pesquera, Mirasso, and Fischer]{largerPhotonicInformationProcessing2012}
L.~Larger, M.~C. Soriano, D.~Brunner, L.~Appeltant, J.~M. Gutierrez,
  L.~Pesquera, C.~R. Mirasso, and I.~Fischer.
\newblock Photonic information processing beyond turing: an optoelectronic
  implementation of reservoir computing.
\newblock 20\penalty0 (3):\penalty0 3241.
\newblock ISSN 1094-4087.
\newblock \doi{10.1364/OE.20.003241}.
\newblock URL \url{https://opg.optica.org/oe/abstract.cfm?uri=oe-20-3-3241}.

\bibitem[Paquot et~al.()Paquot, Duport, Smerieri, Dambre, Schrauwen,
  Haelterman, and Massar]{paquotOptoelectronicReservoirComputing2012}
Y.~Paquot, F.~Duport, A.~Smerieri, J.~Dambre, B.~Schrauwen, M.~Haelterman, and
  S.~Massar.
\newblock Optoelectronic reservoir computing.
\newblock 2\penalty0 (1):\penalty0 287.
\newblock ISSN 2045-2322.
\newblock \doi{10.1038/srep00287}.
\newblock URL \url{https://www.nature.com/articles/srep00287}.

\bibitem[Martinenghi et~al.()Martinenghi, Rybalko, Jacquot, Chembo, and
  Larger]{martinenghiPhotonicNonlinearTransient2012}
Romain Martinenghi, Sergei Rybalko, Maxime Jacquot, Yanne~K. Chembo, and
  Laurent Larger.
\newblock Photonic nonlinear transient computing with multiple-delay wavelength
  dynamics.
\newblock 108\penalty0 (24):\penalty0 244101.
\newblock ISSN 0031-9007, 1079-7114.
\newblock \doi{10.1103/PhysRevLett.108.244101}.
\newblock URL \url{https://link.aps.org/doi/10.1103/PhysRevLett.108.244101}.

\bibitem[Chang and Yeung()]{changRobustPathbasedSpectral2008}
Hong Chang and Dit-Yan Yeung.
\newblock Robust path-based spectral clustering.
\newblock 41\penalty0 (1):\penalty0 191--203.
\newblock ISSN 00313203.
\newblock \doi{10.1016/j.patcog.2007.04.010}.
\newblock URL
  \url{https://linkinghub.elsevier.com/retrieve/pii/S0031320307002038}.

\bibitem[Zhang et~al.()Zhang, Gu, Jiang, Thompson, Cai, Paesani, Santagati,
  Laing, Zhang, Yung, Shi, Muhammad, Lo, Luo, Dong, Kwong, Kwek, and
  Liu]{zhangOpticalNeuralChip2021}
H.~Zhang, M.~Gu, X.~D. Jiang, J.~Thompson, H.~Cai, S.~Paesani, R.~Santagati,
  A.~Laing, Y.~Zhang, M.~H. Yung, Y.~Z. Shi, F.~K. Muhammad, G.~Q. Lo, X.~S.
  Luo, B.~Dong, D.~L. Kwong, L.~C. Kwek, and A.~Q. Liu.
\newblock An optical neural chip for implementing complex-valued neural
  network.
\newblock 12\penalty0 (1):\penalty0 457.
\newblock ISSN 2041-1723.
\newblock \doi{10.1038/s41467-020-20719-7}.
\newblock URL \url{https://www.nature.com/articles/s41467-020-20719-7}.

\bibitem[Song et~al.()Song, Murty~Kottapalli, Goyal, Schölkopf, and
  Fischer]{songLowpowerScalableMultilayer2024}
Alexander Song, Sai~Nikhilesh Murty~Kottapalli, Rahul Goyal, Bernhard
  Schölkopf, and Peer Fischer.
\newblock Low-power scalable multilayer optoelectronic neural networks enabled
  with incoherent light.
\newblock 15\penalty0 (1):\penalty0 10692.
\newblock ISSN 2041-1723.
\newblock \doi{10.1038/s41467-024-55139-4}.
\newblock URL \url{https://www.nature.com/articles/s41467-024-55139-4}.

\bibitem[Hary et~al.()Hary, Brunner, Leybov, Ryczkowski, Dudley, and
  Genty]{haryPrinciplesMetricsExtreme2025}
Mathilde Hary, Daniel Brunner, Lev Leybov, Piotr Ryczkowski, John~M. Dudley,
  and Goëry Genty.
\newblock Principles and metrics of extreme learning machines using a highly
  nonlinear fiber.
\newblock URL \url{http://arxiv.org/abs/2501.05233}.

\bibitem[Castell\'{o}-Lurbe et~al.(2020)Castell\'{o}-Lurbe, Carrascosa,
  Silvestre, D\'{i}ez, Erps, Vermeulen, and Andr\'{e}s]{Castello-Lurbe:20}
David Castell\'{o}-Lurbe, Antonio Carrascosa, Enrique Silvestre, Antonio
  D\'{i}ez, J\"{u}rgen~Van Erps, Nathalie Vermeulen, and Miguel~V. Andr\'{e}s.
\newblock Measurement of the soliton number in guiding media through continuum
  generation.
\newblock \emph{Opt. Lett.}, 45\penalty0 (16):\penalty0 4432--4435, Aug 2020.
\newblock \doi{10.1364/OL.399382}.
\newblock URL \url{https://opg.optica.org/ol/abstract.cfm?URI=ol-45-16-4432}.

\bibitem[Zajnulina()]{zajnulinaShannonEntropyHelps2024}
Marina Zajnulina.
\newblock Shannon entropy helps optimize the performance of a
  frequency-multiplexed extreme learning machine.
\newblock URL \url{http://arxiv.org/abs/2408.07108}.

\bibitem[Rotter()]{rotterConceptFisherInformation2024}
Stefan Rotter.
\newblock The concept of fisher information in scattering problems and neural
  networks.
\newblock In \emph{Nonlinear Optics and its Applications 2024}, volume
  {PC}13004, page PC1300401. {SPIE}.
\newblock \doi{10.1117/12.3022535}.

\bibitem[Dambre et~al.(2012)Dambre, Verstraeten, Schrauwen, and
  Massar]{dambre2012information}
Joni Dambre, David Verstraeten, Benjamin Schrauwen, and Serge Massar.
\newblock Information processing capacity of dynamical systems.
\newblock \emph{Scientific Reports}, 2:\penalty0 514, 2012.
\newblock \doi{10.1038/srep00514}.

\end{thebibliography}

\section{Appendix}
\renewcommand{\thefigure}{\arabic{figure}A} 
\renewcommand{\thetable}{\arabic{table}A}   
\setcounter{figure}{0} 
\setcounter{table}{0}  
\begin{table*}[!h]
\centering
\caption{First MLP, distribution of nodes (1024) across hidden layers, corresponds to the results shown in figure 6}
\label{tab:1A}
\begin{tabular}{p{1.2cm}|p{1.2cm}|p{1.2cm}|p{1.2cm}|p{1.2cm}|p{1.2cm}}
\textbf{Hidden layers} & \textbf{First layer} & \textbf{Second layer} & \textbf{Third layer} & \textbf{Fourth layer} & \textbf{Fifth layer} \\ \hline
1 & 1024 &  &  &  &  \\ 
2 & 512 & 512 &  &  &  \\ 
3 & 512 & 256 & 256 &  &  \\ 
4 & 512 & 256 & 128 & 128 &  \\ 
5 & 512 & 256 & 128 & 64 & 64 \\ 
\end{tabular}
\end{table*}
\begin{table*}[!b]
\centering
\caption{Second MLP, distribution of nodes (400) across hidden layers, corresponds to the results shown in figure 6}
\label{tab: 2A}
\begin{tabular}{p{1.2cm}|p{1.2cm}|p{1.2cm}|p{1.2cm}|p{1.2cm}|p{1.2cm}}
\textbf{Hidden layers} & \textbf{First layer} & \textbf{Second layer} & \textbf{Third layer} & \textbf{Fourth layer} & \textbf{Fifth layer} \\ \hline
1 & 400 &  &  &  &  \\ 
2 & 200 & 200 &  &  &  \\ 
3 & 200 & 100 & 100 &  &  \\ 
4 & 200 & 100 & 50 & 50 &  \\ 
5 & 125 & 100 & 75 & 50 & 50 \\ 
\end{tabular}
\end{table*}
\begin{figure*}[b!]
    \centering
    \includegraphics[width=1\linewidth]{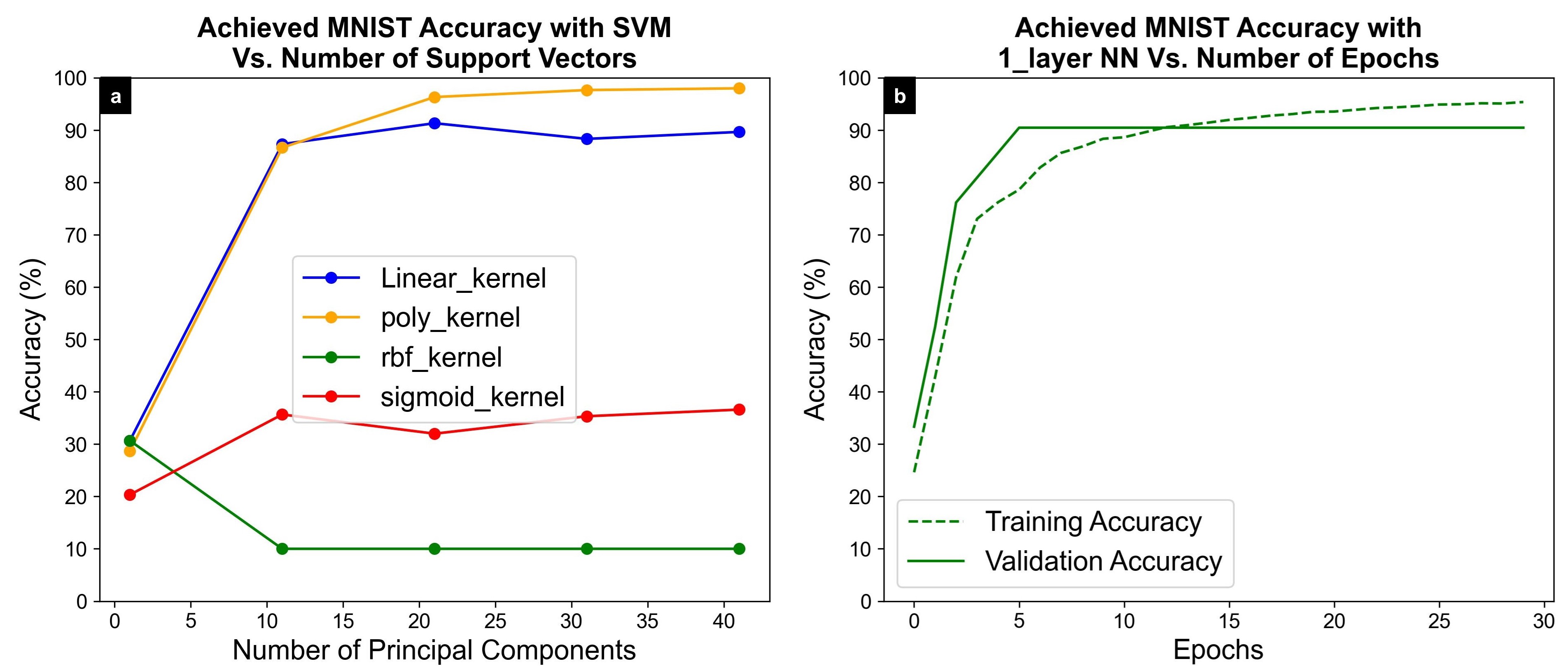}
    \caption{(a) Test accuracy achieved on the MNIST dataset using Support Vector Machines (SVM) with different kernels, plotted against the number of support vectors used. (b) Test accuracy achieved on the MNIST dataset using a one-layer MLP with 10 hidden nodes, evaluated across different numbers of training epochs.}
    \label{fig:1A}
\end{figure*}
\begin{figure*}[!t]
    \centering
    \includegraphics[width=1\linewidth]{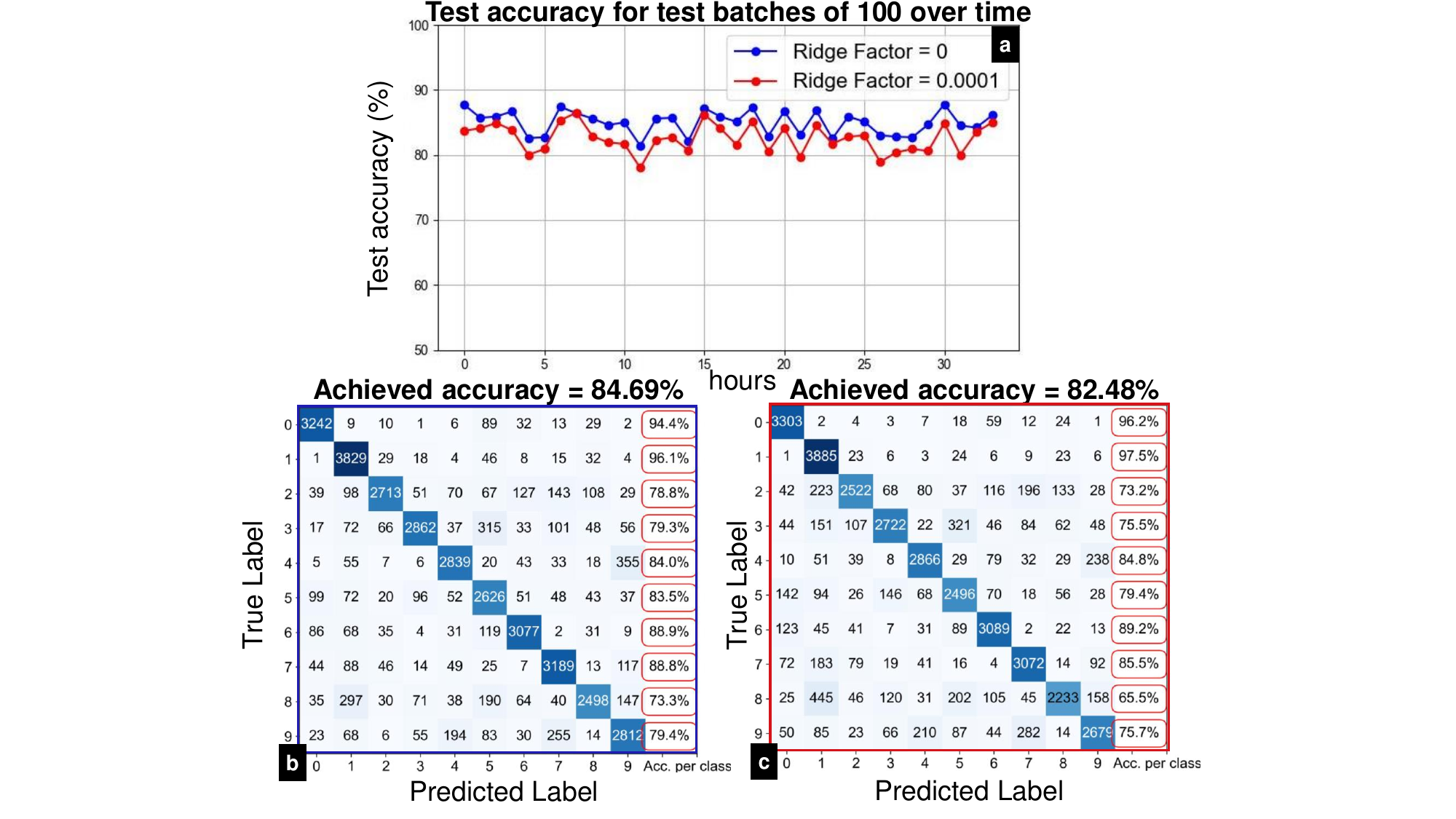}
    \caption{(a) Test accuracy achieved on MNIST samples recorded over a 35-hour period. Each test sample consists of 1000 images, while 3000 images were used for training the system, we trained the system and with and without applying a ridge factor. (b-C) Test-data confusion matrices for both cases.}
    \label{fig:2A}
\end{figure*}
\begin{figure*}[!b]
    \centering
    \includegraphics[width=1\linewidth]{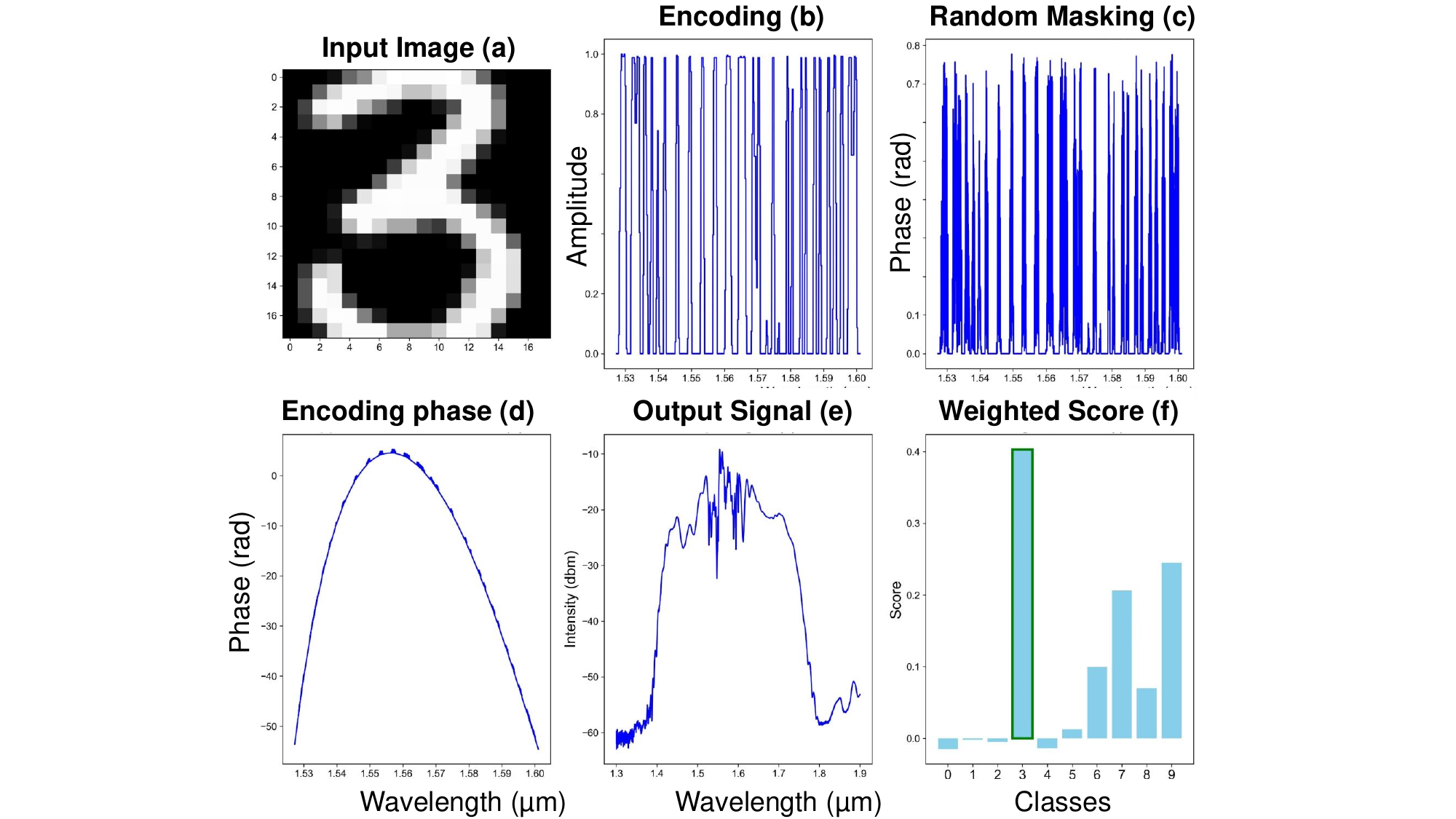}
    \caption{(a) Input MNIST image cropped to an 18×18-pixel window. (b) The corresponding 1D signal derived from flattening the input image. (c) The resulting encoded signal after multiplying the 1D signal by a random signal and the phase scale factor. (d) Encoding phase, generated by adding the masked signal to the optimized zero-phase profile. (e) The corresponding spectral output signal. (f) The weighted output score, obtained by multiplying the read-outs with the trained weight matrix. A winner-takes-all approach is applied, where the predicted class corresponds to the maximum score; in this case, the predicted class is "3".}
    \label{fig:3A}
\end{figure*}
\begin{figure*}[!t]
    \centering
    \includegraphics[width=0.9\linewidth]{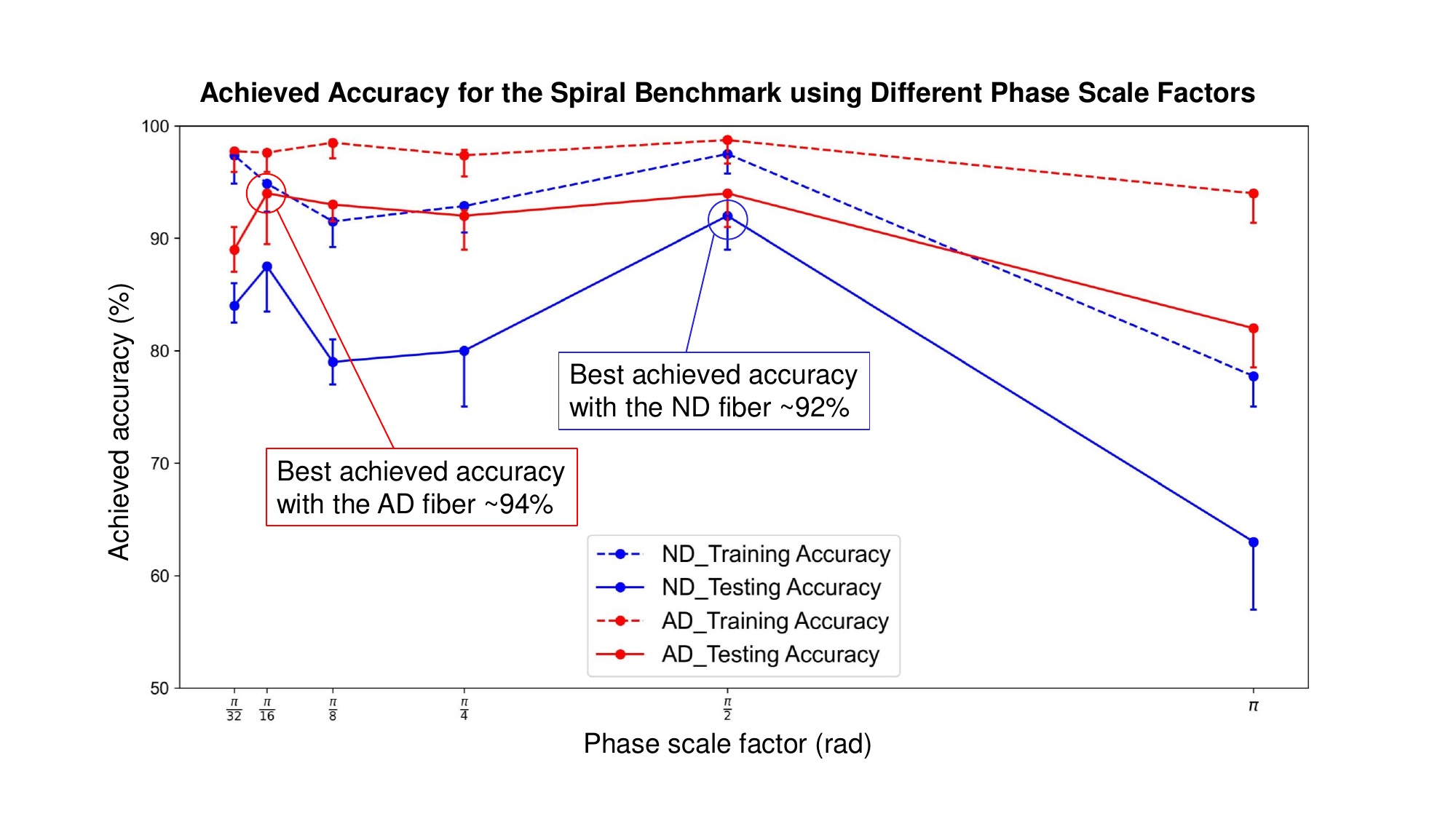}
    \caption{The achieved accuracy for the spiral benchmark with $\theta_{max}=10\pi$, evaluated using different phase scale factors. All results were recorded under consistent conditions, including zero attenuation (0 dB) and identical training and testing samples.}
    \label{fig:4A}
\end{figure*}
\end{document}